\documentclass[prl,twocolumn,draft,amsmath,showpacs]{revtex4}
\usepackage{graphics}\input epsf

\begin{document}

\title{Hole stripe and orbital fluctuating state in LaMnO$_{3+\delta}$
($0.085\leq \delta\leq0.125$) unveiled by Raman spectroscopy}

\author{Yu. G. Pashkevich,$^1$ V. P. Gnezdilov,$^2$
  P. Lemmens,$^{3}$ K.-Y. Choi,$^{4, \dag}$  G. G\"{u}ntherodt,$^5$ A. V. Eremenko,$^2$
  \\ D. Nabok,$^1$ V. I. Kamenev,$^1$ S. N. Barilo,$^6$
  S. V. Shiryaev,$^6$ and  A. G. Soldatov$^6$}

\affiliation{$^1$ A. A. Galkin Donetsk Phystech NASU, 83114
Donetsk, Ukraine}

\affiliation{$^2$ B. I. Verkin Institute for Low Temperature
Physics NASU, 61164 Kharkov, Ukraine}

\affiliation{$^3$ Institute for Physics of Condensed Matter, TU
Braunschweig, D-38106 Braunschweig, Germany}

\affiliation{$^4$ Institute for Materials Research, Tohoku
University, Katahira 2-1-1, Sendai 980-8577, Japan}

\affiliation{$^5$ 2. Physikalisches Institut, RWTH Aachen, 52056
Aachen, Germany}

\affiliation{$^6$ Institute of Physics of Solids \&
Semiconductors, Academy of Sciences, 220072 Minsk, Belarus}

\date{\today}


\begin{abstract}
Giant softening by 30~cm$^{-1}$ of the 490- and 620-cm$^{-1}$
modes is observed by Raman scattering measurements below the Curie
temperature of single crystalline LaMnO$_{3+\delta}$ ($0.085\leq
\delta\leq0.125$). A pseudogap-like electronic continuum and a
Fano antiresonance at 144 cm$^{-1}$ appear below  the charge
ordering temperature. This gives evidence for the presence of an
orbital fluctuating state and the formation of a hole stripe,
respectively. This is further corroborated by a unstructured
broadening and shifting of multiphonon features with increasing
doping $\delta$. Our study suggests the significance of double
exchange mechanism in the charged ordered insulating state.
\end{abstract}


\maketitle



One topical issue in the physics of manganites is on the role of
orbital degrees of freedom as well as the state of an orbital
subsystem at different doping level in explaining a complex phase
diagram \cite{Tokura,Brink-review}. The most intriguing feature is
found in the low-doped region  La$_{1-x}$Sr$_x$MnO$_3$
($x=0.11-0.17$), which shows the unusual coexistence of
ferromagnetism and insulating behavior
\cite{Endo99,Dabrowski99,niemoeller,Yama96,Yama00,klingeler,Geck03,Geck04,choi05}.

To account for the occurrence of a ferromagnetic insulating state
several different models have been proposed on the orbital and
charge ordering pattern. There is some evidence for a structural
modulation with alternating hole-poor and hole-rich planes along
the $c$ axis \cite{Yama96,Yama00,Mizokawa1} as well as for a
formation of orbital polarons
\cite{Geck03,Geck04,choi05,Kilian99}. However, a complete picture
is still lacking.

In this paper,  we report Raman scattering measurements on the
lightly oxygen-doped manganites LaMnO$_{3+\delta}$. We provide
evidence for the formation of a hole stripe seen from a
pseudogap-like electronic response below the charge ordering
temperature. Furthermore, a giant and continuous softening of Mn-O
bond stretching modes below the Curie temperature and a
unstructured, broadened multiphonon feature suggest a fluctuating
orbital state in ferromagnetic samples ($0.085\leq
\delta\leq0.125$). This together with a Fano antiresonance at 144
cm$^{-1}$ signals the significance of double exchange interaction
in the ferromagnetic insulating state.

LaMnO$_{3+\delta}$ single crystals were grown by using a modified
method of McCarrol {\it et al.} \cite{McCar}, which  allows for
doping without inducing a large cation mass difference. The true
crystallographic formula corresponds to La$_{1-x}$Mn$_{1-y}$O$_3$
with $3/2(x+y)\approx\delta$. Cation vacancies were controlled by
varying the melting temperature. Samples were characterized by
X-ray diffraction, magnetic susceptibility and chemical analysis
\cite{Barilo}. The structural and magnetic properties are
summarized in Table I. Our results are consistent with the
previous ones \cite{Ritter,Prado}. Furthermore, the magnetic and
transport behaviors are quite similar to the lightly-doped
La$_{1-x}$Sr$_x$MnO$_3$ ($0.11\leq x\leq0.15$) (see Refs.
\cite{Geck04,Barilo}). The slight discrepancy, for example, seen
in the irreversible magnetization between field and zero-field
cooling should be attributed to additional La-cation disorders.

Raman scattering experiments were performed using the excitation
line $\lambda= 514.5$~nm of an Ar$^{+}$ laser in a
quasibackscattering geometry. The laser power of 5~mW was focused
to a 0.1~mm diameter spot on the (010) surface.  The scattered
spectra were collected by a DILOR-XY triple spectrometer and a
nitrogen cooled CCD detector with a spectral resolution of $\sim
1$ cm$^{-1}$.

\begin{table}
\caption{Structural and magnetic properties of LaMnO$_{3+\delta}$
with O $(c\leq b/\surd 2<a)$ and O' $(b/\surd 2<c<a)$ orthorhombic
phases with Pnma space group symmetry and R a rhombohedral phase
with R$\bar3$c space group symmetry.} \label{TablI}
  \centering
\begin{tabular}{ccccc}
\\
\hline \hline \,\,sample\,\, & \,\,magnetic\,\,  &
\,\,critical\,\, & \,\,V/f.u. (\AA$^3$)\,\, &
crystal\\
$\rm 3+\delta$ &ordering  & temperature & T=300 K  & structure
\\ \hline
3.071 & AFM & $T\rm _N$= 128~K  & 59.73 & O' \\

3.085 & FM & $T\rm _C$= 148~K  & 59.54 & O' \\

3.092 & FM & $T\rm _C$= 178~K  & 59.45 & O \\

3.096 & FM & $T\rm _C$= 186~K  & 59.41 & O or R \\

3.125 & FM & $T\rm _C$= 248~K  & 59.06 & R \\
\hline
\end{tabular}

\end{table}

\begin{figure}[th]
      \begin{center}
       \leavevmode
       \epsfxsize=8.5cm \epsfbox{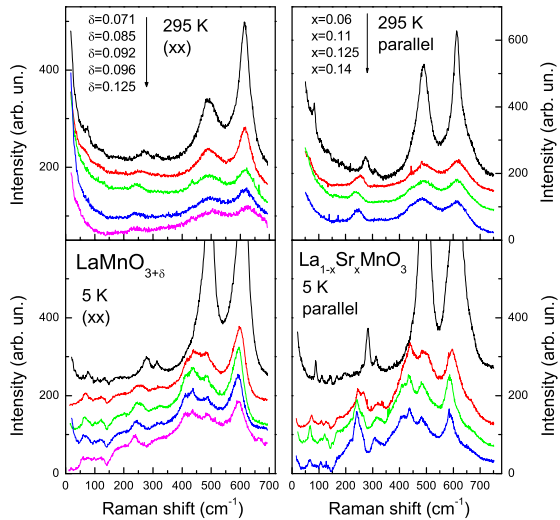}
        \caption{(online color) (Left panels) Raman spectra of LaMnO$\rm _{3+\delta}$
        ($\delta$=0.071, 0.085, 0.092,
0.096, and 0.125) in ($xx$) polarization as a function of doping
at 5 and 295 K, respectively. (Right panels) For comparison, Raman
spectra of La$_{1-x}$Sr$_x$MnO$_3$ ($x$=0.06, 0.11, 0.125, and
0.14) are presented together at the respective temperature
\cite{choi05}.} \label{Fig1}
\end{center}
\end{figure}

Figure~1 displays  the doping dependence of polarized Raman
spectra of LaMnO$_{3+\delta}$  ($\delta$=0.071, 0.085, 0.092,
0.096, and 0.125) at 5 and 295 K, respectively. They are compared
to the lightly doped manganites La$_{1-x}$Sr$_x$MnO$_3$ ($x$=0.06,
0.11, 0.125, and 0.14) that are a counterpart of the title
compound \cite{choi05}. Remarkably, the observed Raman spectra are
parallel to each other. For the antiferromagnetic (AF) sample
($\delta$=0.071 and $x$=0.06) weak peaks arising from vibrations
of (La/Sr) cations and rotations of the MnO$_6$ octahedra can be
resolvable in addition to the three main peaks; the out-of-phase
rotational mode around 250 cm$^{-1}$, the Jahn-Teller (JT) mode
around 490 cm$^{-1}$, and the breathing mode around 620 cm$^{-1}$.
At low temperatures the ferromagnetic insulating (FMI) samples
($\delta=0.085 - 0.125$ and $x=0.11 - 0.14$) show commonly the
extra phonon peaks which are split off from the high-temperature
modes. This is due to activated modes induced by the charge and
orbital ordering. Noticeably, all FMI samples have the identical
Raman spectra, irrespective of doping and compound. This implies
that all FMI samples are characterized by the same charge and
orbital ordering structure in spite of additional, disordered
holes as well as of chemical disorders. X-ray measurements of
La$_{1-x}$Sr$_x$MnO$_3$ support further this, which exhibits the
same superstructure reflections of nearly equal scattering
intensity for all $x$~\cite{niemoeller}. This might be related to
a robustness of the specific orbital and charge order found at
$x=1/8$.

\begin{figure}[th]
      \begin{center}
       \leavevmode
       \epsfxsize=8cm \epsfbox{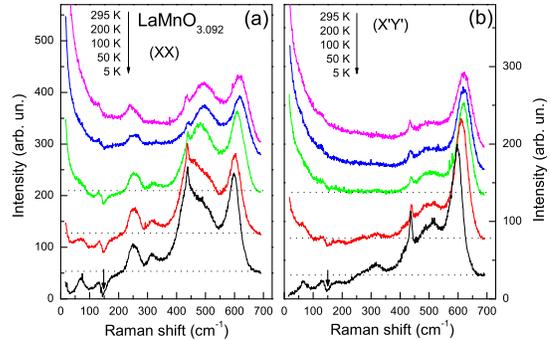}
        \caption{(color online) Temperature-dependence of Raman spectra
         of LaMnO$\rm _{3+\delta}$ ($\delta=0.092$) in (a) ($xx$) and (b) ($x'y'$) polarizations.
          The horizontal dotted lines are a guide for
         an electronic background. The vertical arrows indicate antiresonance
         at 144 cm$^{-1}$. } \label{F2}
\end{center}
\end{figure}

Shown in Figs. 2 is the temperature dependence of the Raman
spectra of the representative sample at $\delta=0.092$ in ($xx$)
and ($x'y'$) polarizations.  They exhibit intriguing features.
With decreasing temperature the Mn-O stretching modes undergo a
large softening in addition to the appearance of new phonon peaks.
Furthermore, the low-frequency quasielastic response is
systematically suppressed while a pseudogap opens at low
temperatures. Noticeably, a strong Fano antiresonance is observed
at 144 cm$^{-1}$ which corresponds to the rotation of the MnO$_6$
octahedra.

\begin{figure}[th]
      \begin{center}
       \leavevmode
       \epsfxsize=8cm \epsfbox{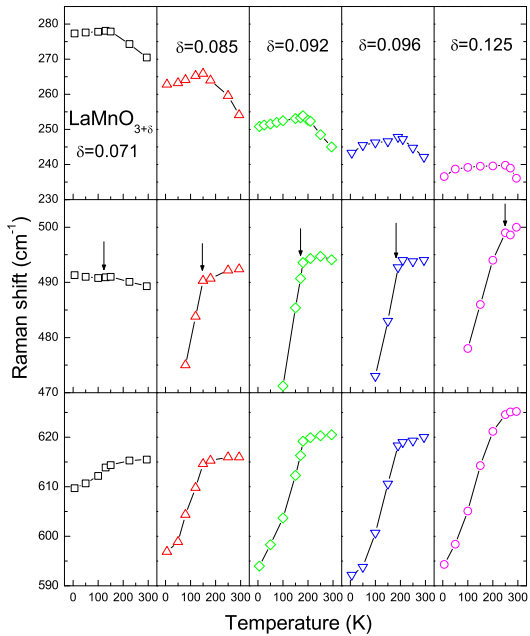}
        \caption{(color online) Temperature dependence of peak
position of the 250-, 490-, and 620-cm$\rm ^{-1}$ mode as a
function of doping $\delta$. The vertical arrows indicate the
magnetic ordering temperature. } \label{Fig5}
\end{center}
\end{figure}

The most salient feature can be seen in the temperature dependence
of the main peaks at 250, 490, and 620 cm$\rm ^{-1}$ as summarized
in Fig. 3 as a function of $\delta$. The experimental spectra are
analyzed by a sum of a Lorentzian profile. Error bars are of the
symbol size.

First, let us begin with the out-of-phase rotational mode around
250 cm$\rm ^{-1}$. This mode is associated with the tolerance
factor of the manganites. That is, it provides direct information
about octahedral tiltings and Mn-O-Mn bond angles. With increasing
$\delta$ the mode softens by 40 cm$^{-1}$. The large doping
dependence should be attributed to the suppression of the static
JT distortion as hole mobilities increase. This is also reflected
in the structural transition from orthorhombic to rhombohedral
phase (see Table I). Further, upon
 cooling the 250-cm$\rm^{-1}$ mode first hardens up to T$\rm_C$ and then softens
slightly. The magnitude of the hardening decreases as $\delta$
increases, for example, from 12 cm$\rm ^{-1}$ at $\delta=0.085$ to
4 cm$\rm ^{-1}$ at $\delta=0.125$. Here note that between T$_{JT}$
and T$_C$ there exits a $d_{3x^2-r^2}$/$d_{3y^2-r^2}$ orbital
ordering which is similar to that of LaMnO$_3$
\cite{Geck03,Geck04}. This antiferro-orbital ordering is
stabilized by the cooperative JT distortion.  Therefore, the
decrease of the hardening with increasing $\delta$ further
evidences the reduction of the LaMnO$_3$-type orbital order. In
this respect, moreover, the softening by 2-3 cm$\rm ^{-1}$ below
T$_{C}$ can be interpreted in terms of the fact that the
LaMnO$_3$-type orbital above T$_C$ evolves to the different type
of orbital below T$_C$. Actually, resonant X-ray scattering
measurements unveil the rearrangement of the orbital ordering
through T$_{C}$ \cite{Geck03,Geck04}.

Next, we will address the temperature dependence of the MnO$_6$
vibrational modes. Upon cooling the JT mode of the AF sample at
about 490 cm$^{-1}$ shows a slight hardening by 3 cm$^{-1}$ while
the breathing mode at about 620 cm$^{-1}$ undergoes a moderate
softening by 6 cm$\rm ^{-1}$. For the FMI samples the respective
modes turn to a giant softening by 20-30~cm$\rm ^{-1}$ below T$\rm
_C$. As $\delta$ increases, the softening becomes enhanced.
Furthermore, the softening does not exhibit any saturation even at
very low temperatures. It is worth to note that such a giant
softening is exclusively seen at the breathing mode for the
lightly doped La$_{1-x}$Sr$_x$MnO$_3$ \cite{choi05}. This is
ascribed to the intrinsic coupling of the breathing mode to
orbital polarons and is taken as evidence for the formation of
orbital polarons in the hole-rich sites. Most probably, the
hole-poor planes are composed of a LaMnO$_3$-type orbital
~\cite{Mizokawa1}. In this case, the La-site disorder in LaMnO$\rm
_{3+\delta}$  can lead to the additional softening of the JT-mode
in contrast to La$_{1-x}$Sr$_x$MnO$_3$. Therefore, we conclude
that the hole-poor planes contain a LaMnO$_3$-type orbital while
the hole-rich planes are dominated by orbital polarons. Moreover,
the softening without saturation indicates a fluctuating nature of
the underlying orbital state.

\begin{figure}[th]
      \begin{center}
       \leavevmode
       \epsfxsize=6cm \epsfbox{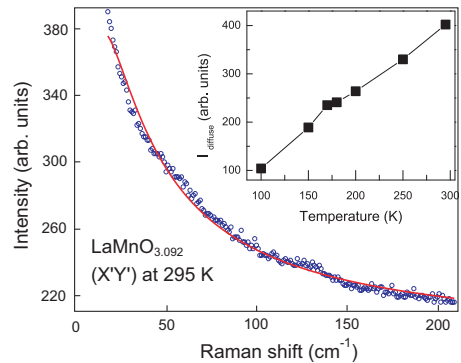}
        \caption{(color online) A fitting of the low-frequency electronic
         response of $\delta=0.092$ at 295 K using Eq. (1). Inset: Temperature dependence
        of the scattering amplitude, I$_{diffuse}$. } \label{Fig4}
\end{center}
\end{figure}

We will now examine the electronic response to get information
about hole dynamics.  At high temperatures we have observed
pronounced quasielastic Raman response. The observed low-frequency
response is well described by a diffusive scattering
\cite{choi03};

\begin{equation}
I(\omega, T)= \frac{1}{1-exp(\hbar\omega/kT)}\times
\frac{1_{diffuse}\omega\Gamma}{\omega^2 +\Gamma^2},
\end{equation}
where the first term is the Bose-thermal factor and $I_{diffuse}$
is the scattering amplitude and  $\Gamma$ the scattering rate.
Figure 4 displays the typical fit using Eq. (1) at 295 K. The
temperature dependence of the scattering amplitude is shown in the
inset of Fig. 4. The scattering rate of $\Gamma=13.5-14.5$
cm$^{-1}$ is nearly temperature independent. Its magnitude is two
orders of magnitude smaller than that of the optimal doped
La$_{1-x}$Sr$_x$MnO$_3$ \cite{choi03}.  The scattering amplitude
of I$_{diffuse}$ decreases linearly upon cooling. It persists much
below the charge ordering temperature, T$\rm _{CO}$. The diffusive
scattering originates from fluctuations related to electronic
scattering from spins and impurities.  Therefore, this suggests
that holes are not totally frozen-in below T$\rm _{CO}$. At low
temperatures  the depletion of spectral weight is observed below
250 cm$^{-1}$ (see Figs. 2). The presence of a pseudogap-like
behavior below the charge ordering temperature is due to the
formation of a hole stripe as in La$_{2-x}$Sr$_x$NiO$_4$
\cite{Vladimir}. Moreover, a Fano antiresonance appears at 144
cm$^{-1}$, pointing to substantial electron-phonon coupling. Here
we will remind that a stripe is formed to gain a kinetic energy of
holes in the AF background in the cuprates and nickelates. In
analogy, a hole stripe in the manganites arises from maximizing
locally the gain of double exchange energy while keeping globally
an insulating behavior. The evolution of LaMnO$_3$-type orbitals
to the new orbital state including orbital polarons through T$_C$
is also initiated by double-exchange mechanism \cite{Geck04}.

\begin{figure}[th]
      \begin{center}
       \leavevmode
       \epsfxsize=8cm \epsfbox{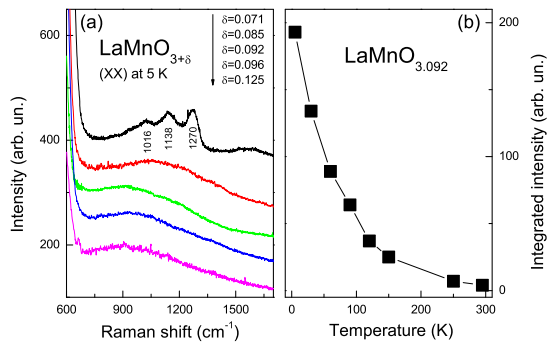}
        \caption{(a) Raman spectra at T=5~K in the orbiton frequency regime.
(b) Temperature evolution of integrated intensity of the maximum
at 900~cm$\rm ^{-1}$ in sample M3.} \label{Fig6}
\end{center}
\end{figure}

Finally, we will turn to the high-frequency Raman spectra of
LaMnO$\rm _{3+\delta}$ at 5~K which are displayed in Fig. 5(a) as
a function of $\delta$. The AF sample shows similar results as
La$_{1-x}$Sr$_x$MnO$_3$ with respect to frequency, number of
modes, and temperature dependence \cite{KY05}. For the FMI sample
of $\delta=0.085$ the three peaks coalesce into a broad maximum
near 1100~cm$\rm ^{-1}$. Upon further increasing $\delta$ the
broad maximum is shifted to about 900~cm$\rm ^{-1}$. One-phonon
scattering of the 490 and 620~cm$\rm ^{-1}$ modes softens by
maximum 5\% in going from the $\delta=0.085$ to $\delta=0.125$
sample. In contrast, the spectral weight of the broad maximum
softens by 20\% for the respective samples. Such a behavior can
definitely not be understood within pure multiphonon processes.
Note that La$_{1-x}$Sr$_x$MnO$_3$ shows also a broadened maximum
around 1000~cm$\rm ^{-1}$ at $x$=0.125 \cite{KY05}. This rules out
chemical disorders as a plausible origin. The temperature
dependence of the integrated intensity of the maximum at
$\delta=0.092$ is given in Fig. 5(b). With decreasing temperature
the intensity exhibits the 1/T-divergence. This suggests that the
observed maximum is closely related to the orbital rearrangement
through T$_C$.

In the AF phase resonant Raman scattering measurements unveiled
that multiphonon scattering is govern by resonant process induced
by an orbital flip of a $d_{3x^2-r^2}$/$d_{3y^2-r^2}$ orbital
ordered state~\cite{Krueger}. In this situation, a detailed
feature of the multi-phonon scattering relies on the underlying
orbital ordering pattern. Upon going through the AF/FMI boundary
the $d_{3x^2-r^2}$/$d_{3y^2-r^2}$ orbitals are rearranged to the
alternating orbitals consisting of orbital polarons in the
hole-rich planes and a $d_{3x^2-r^2}$/$d_{3y^2-r^2}$ orbital in
the hole-poor planes. Thus, orbital excitons, which are
responsible for the three-peaks feature in the AF phase, will
smear out in the FMI phase. This naturally explains a shift of the
multiphonon scattering to lower energy. Further, the unstructured
maximum is associated with a fluctuating orbital state. This
interpretation of the high-frequency maxima is consistent with the
observed giant softening in terms of a mutual coupling between
phonons and the orbital fluctuating state.

In summary, our Raman study of the lightly doped manganites $\rm
LaMnO_{3+\delta}$ reveals the formation of a hole-stripe and an
orbital fluctuating state in the FMI phase. A Fano antiresonance
gives evidence that this is owed to the gain of double-exchange
energy in the insulating background. This demonstrates the
importance of electronic correlation effects in the manganites.

We thank J. van den Brink, B. B\"{u}chner, M. Braden, G.
Khaliulin, D. Khomskii, M. Mostovoy and M. M. Savosta for useful
discussions. This work was supported in part by the NATO
Collaborative Linkage Grant PST.CLG.977766.

$^\dag$To whom correspondence should be addressed. E-mail:
choi@imr.tohoku.ac.jp.

\end{document}